\documentclass[structabstract]{aa}  

\usepackage{graphicx}
\usepackage{txfonts}
\usepackage{multirow}

\begin{document}

   \title{The two-component jet of GRB 080413B}

   \author{R. Filgas\inst{1} \and T. Kr\"uhler\inst{1,2} \and J. Greiner\inst{1} \and 
           A. Rau\inst{1} \and E. Palazzi\inst{3} \and S. Klose\inst{4} \and
           P. Schady\inst{1} \and A. Rossi\inst{4} \and 
           P. M. J. Afonso\inst{1} \and L. A. Antonelli\inst{5} \and C. Clemens\inst{1} 
           \and S. Covino\inst{6} \and P. D'Avanzo\inst{6} 
           \and A. K\"upc\"u Yolda\c{s}\inst{7,8} \and M. Nardini\inst{1} \and A. Nicuesa Guelbenzu\inst{4} \and
           F. Olivares E.\inst{1} \and A. C. Updike\inst{9} \and A. Yolda\c{s}\inst{1,8}}

   \institute{Max-Planck-Institut f\"ur extraterrestrische Physik,
     Giessenbachstra\ss{}e 1, D-85748 Garching, Germany, \\     
   \email{filgas@mpe.mpg.de}
        \and
     Universe Cluster, Technische Universit\"at M\"unchen, 
     Boltzmannstra\ss{}e 2, D-85748 Garching, Germany
        \and
     INAF - IASF di Bologna, via Gobetti 101, I-40129, Bologna, Italy     
        \and
     Th\"uringer Landessternwarte Tautenburg, Sternwarte 5,
     D-07778 Tautenburg, Germany
         \and
     INAF - Osservatorio Astronomico di Roma, via di Frascati 33, I-00040, Monteporzio Catone (Roma), Italy
         \and
     INAF - Osservatorio Astronomico di Brera, via Bianchi 46, I-23807, Merate, Italy
         \and
     European Southern Observatory, Karl-Schwarzschild-Stra\ss{}e 2, 
     D-85748 Garching, Germany
         \and
     Institute of Astronomy, University of Cambridge, Madingley Road CB3 0HA, Cambridge, UK
         \and
     Department of Physics and Astronomy, Clemson University, Clemson, 
     SC 29634-0978, United States of America}

   \date{Received 01 September 2010 / Accepted 02 October 2010}

  \abstract
   {}
   {The quick and precise localization of GRBs by the \emph{Swift}
    telescope allows the early evolution of the afterglow light curve to be 
    captured by ground-based telescopes. With GROND measurements we can 
    investigate the optical/near-infrared light curve of the afterglow of
    gamma-ray burst 080413B in the context of late rebrightening.}
   {Multi-wavelength follow-up observations were performed on the afterglow
    of GRB 080413B. X-ray emission was detected by the X-ray telescope onboard
    the \emph{Swift} satellite and obtained from the public archive. 
    Optical and near-infrared photometry was performed 
    with the seven-channel imager GROND mounted at the MPG/ESO 2.2~m telescope and 
    additionally with the REM telescope, both in La Silla, Chile. The light curve model 
    was constructed using the obtained broad-band data.}
   {The broad-band light curve of the afterglow of GRB 080413B is well fitted with an on-axis 
    two-component jet model. The narrow ultra-relativistic jet is responsible for the initial decay, 
    while the rise of the moderately relativistic wider jet near its deceleration time is the cause
    of the rebrightening of the light curve. The later evolution of the optical/NIR light curve
    is then dominated by the wide component, the signature of which is almost negligible in the 
    X-ray wavelengths. These components have opening angles of $\theta_n \sim 1.7^\circ$
    and $\theta_w \sim 9^\circ$, and Lorentz factors of $\Gamma_n > 188$ and $\Gamma_w \sim 18.5$.
    We calculated the beaming-corrected energy release to be $E_{\gamma} = 7.9 \times 10^{48}$~erg.}
   {}
   \maketitle 
   
\section{Introduction}
   Gamma-ray burst (GRB) afterglows are commonly interpreted in the framework of the 
   standard synchrotron shock model, in which an ultra-relativistic shock is expanding into 
   the ambient medium swept up by the blast wave (M\'esz\'aros \cite{meszaros2}; Zhang \& M\'esz\'aros 
   \cite{zhang3}; Piran \cite{piran}). For the simplified assumption that the shock front is spherical and 
   homogeneous, a smooth afterglow light curve is expected. 
   This smooth power-law decay with time was a common phenomenon in most
   of the pre-\emph{Swift} GRBs (Laursen \& Stanek \cite{laursen}), because 
   the afterglow observations typically began $\sim$1 day after the burst
   compared to now when we can be on-target within minutes.
   
   The \emph{Swift} satellite (Gehrels et al. \cite{gehrels}) 
   allows studies of the early afterglow phase thanks to its rapid slew, 
   a precise localization
   of GRBs with its Burst Alert Telescope (BAT, Barthelmy et al. \cite{barthelmy}),
   and the early follow-up with two telescopes sensitive at X-ray (XRT, Burrows et al. \cite{burrows})
   and ultraviolet/optical (UVOT, Roming et al. \cite{roming}) 
   wavelengths. Since its launch in 2004, \emph{Swift}, together with ground-based
   follow-up telescopes, has provided many early and well-sampled afterglow light curves
   deviating from the smooth power-law decay (Panaitescu et al. \cite{panaitescu3};
   Nousek et al. \cite{nousek}; Zhang et al. \cite{zhang2}; Panaitescu et al. \cite{panaitescu4}). 
   Such variability can shed light on the central engine and its surroundings.
   
   Several major scenarios have been proposed for afterglow variability. The
   reverse shock emission might add to the emission from the forward shock 
   (see \textsection 4.1, Sari \& Piran \cite{sari2}; M\'esz\'aros 
   \& Rees \cite{meszaros}; Zhang et al. \cite{zhang}; Kobayashi \& Zhang \cite{kobayashi}), 
   the shock might be refreshed by slower shells catching
   up with the decelerating front shells (see \textsection 4.2, Rees \& M\'esz\'aros \cite{rees}; 
   Panaitescu \cite{panaitescu2}; Sari \& M\'esz\'aros \cite{sari3}; Panaitescu et al.\cite{panaitescu}; 
   Granot et al. \cite{granot2}; Kumar \& Piran \cite{kumar}), the ambient density 
   profile into which the blast wave expands might not be homogeneous (see
   \textsection 4.3, Lazzati et al. \cite{lazzati}; Nakar et al. \cite{nakar2}; 
   Zhang et al. \cite{zhang2}; Nakar \& Piran \cite{nakar3}; Ioka et al. \cite{ioka}; 
   Wang \& Loeb \cite{wang}; Dai \& Lu \cite{dai},;Nakar \& Granot \cite{nakar}), or the 
   jet may have an angular structure different from a 
   top hat (see \textsection 4.4, Peng et al. \cite{peng}; Granot et al. \cite{granot}; 
   Berger et al. \cite{berger}; Racusin et al. \cite{racusin2}).
   
   Here we provide details of the \emph{Swift}, GROND, and REM observations of the 
   afterglow of GRB 080413B and test
   the above alternative scenarios for consistency with these data. 
   Throughout the paper, we adopt the convention that the flux density 
   of the GRB afterglow can be described as $F_\nu (t) \propto t^{\alpha} \nu^{-\beta}$.
   
\section{Observations}

\subsection{Swift}
   The \emph{Swift}/BAT triggered by the long GRB 080413B at $T_0 =$ 08:51:12 UT started 
   slewing to the burst after 70~seconds (Stamatikos et al. \cite{stamatikos}). 
   The mask-weighted light curve shows a single FRED-like peak starting 
   at $T_0-1.1$~s, peaking at $T_0+0.2$~s, and returning to baseline at $\sim T_0+30$~s.
   The measured $T_{90}$ (15-350 keV) is $8.0 \pm 1.0$~s (Barthelmy et al. \cite{barthelmy2}).
   The BAT prompt emission spectrum was fitted using the Band function with a photon index 
   of $\alpha=-1.24\pm0.26$ and an $E_{\rm peak}=67^{+13}_{-8}$~keV (Krimm et al. \cite{krimm}). 
   By integrating the GRB spectrum using the Band function, we estimate the event fluence
   in the 15-150 keV energy range to be $3.1\pm0.12 \times 10^{-6}$~erg/cm$^2$ 
   (Krimm et al. \cite{krimm}). With a standard concordance cosmology ($H_0 = 71.0$~km/s/Mpc, 
   $\Omega_M$ = 0.27, $\Omega_{\Lambda} = 0.73$, Komatsu et al. \cite{komatsu}), and a redshift 
   of $z=1.1$ (Fynbo et al. \cite{fynbo}), the bolometric (1keV - 10MeV) energy release of
   GRB~080413B is $E_{\rm iso} = 1.8 \times 10^{52}$~erg, with a rest-frame $E_{\rm peak}$ of $\sim$150~keV.
   The difference between this value and the value in Krimm et al. (\cite{krimm}) is only due to different
   set of cosmological parameters used.

   The \emph{Swift}/XRT started observations of the 
   field of GRB~080413B 131.2~s after the trigger (Stamatikos et al. \cite{stamatikos}; Troja 
   \& Stamatikos \cite{troja}). XRT data 
   were obtained from the public \textit{Swift} archive and reduced in the standard manner using the xrtpipeline 
   task from the HEAsoft package, with response matrices from the most recent CALDB release. The XRT light 
   curve was obtained from the XRT light curve repository (Evans et al. \cite{evans}, Evans et al. \cite{evans2}). 
   Spectra were grouped using the grppha task and fitted with the GROND data in XSPEC v12 using 
   $\chi^2$ statistics. The combined optical/X-ray spectral energy distributions were fitted with power-law 
   and broken power-law models and two absorbing columns: one Galactic foreground with a hydrogen 
   column of $N_H = 3.1\times10^{20}$~cm$^{-2}$ (Kalberla et al. \cite{kalberla}) and another one that is 
   local to the GRB 
   host galaxy at $z=1.1$. Only the latter was allowed to vary in the fits. To investigate the dust 
   reddening in the GRB environment, the zdust model was used, which contains Large and Small Magellanic Clouds 
   (LMC, SMC) and Milky Way (MW) extinction laws from Pei (\cite{pei}).

\subsection{REM}
   The Rapid Eye Mount (REM, Zerbi et al. \cite{zerbi}) 60~cm robotic telescope, located at 
   the ESO La Silla observatory (Chile), reacted promptly and began observing 
   GRB~080413B on April 13 08:52:13 UT, about 76 s after the GRB trigger time. A 
   transient source was detected both in the $R$ and $H$ bands, and follow--up 
   observations lasted for $\sim 1$~hr. 
   The afterglow is well detected only up to about 300 s, then its brightness 
   falls below the instrument detection limits in both filters.

   Each single $H$-band observation was performed with a dithering sequence of five images 
   shifted by a few arcsec. These images are automatically elaborated using the jitter 
   script of the eclipse (Devillard \cite{devillard}) package. The script aligns the images and co-adds all the 
   frames to obtain one average image for each sequence. The $R$-band images were reduced 
   using standard procedures. A combination of the 
   IRAF\footnote{IRAF is the Image Reduction and Analysis Facility made available to 
   the astronomical community by the National Optical Astronomy Observatories, which 
   are operated by AURA, Inc., under contract with the U.S. National Science 
   Foundation. It is available at {\tt http://iraf.noao.edu/}}, and Sextractor 
   packages (Bertin \& Arnouts \cite{bertin}) were then used to perform aperture 
   photometry.

   The photometric calibration for the $H$ band was accomplished by applying average 
   magnitude shifts to the ones of bright, isolated, unsaturated stars in the field, 
   as reported in the 2MASS catalog. The optical data were calibrated using 
   instrumental zero points, checked with observations of standard stars in the SA96 
   Landolt field (Landolt \cite{landolt}). All data were then cross-calibrated using
   GROND photometry to obtain consistent results.

\begin{figure}[h]
   \resizebox{\hsize}{!}{\includegraphics{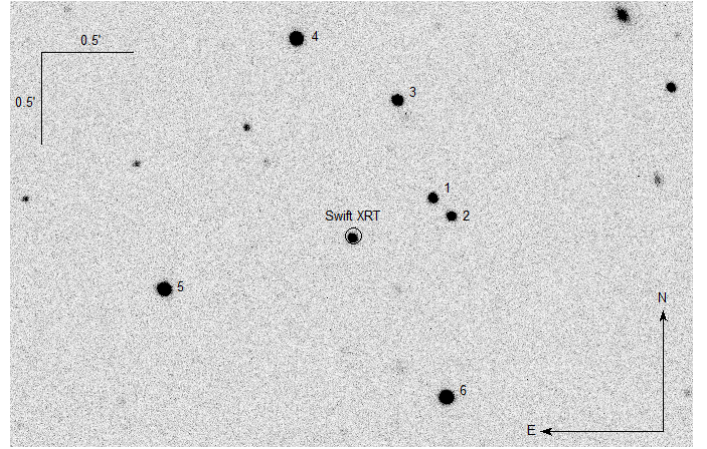}}
   \caption{GROND $r'$ band image of the field of GRB 080413B obtained  
              342~s after T$_0$. The optical afterglow is shown inside the \emph{Swift} XRT
               error circle. The secondary standard stars are numbered from $1$ to $6$ and
             their magnitudes reported in Table \ref{standards}.}
   \label{chart}
\end{figure} 

\begin{figure*}[ht]
   \centering
   \sidecaption
   \includegraphics[width=12cm]{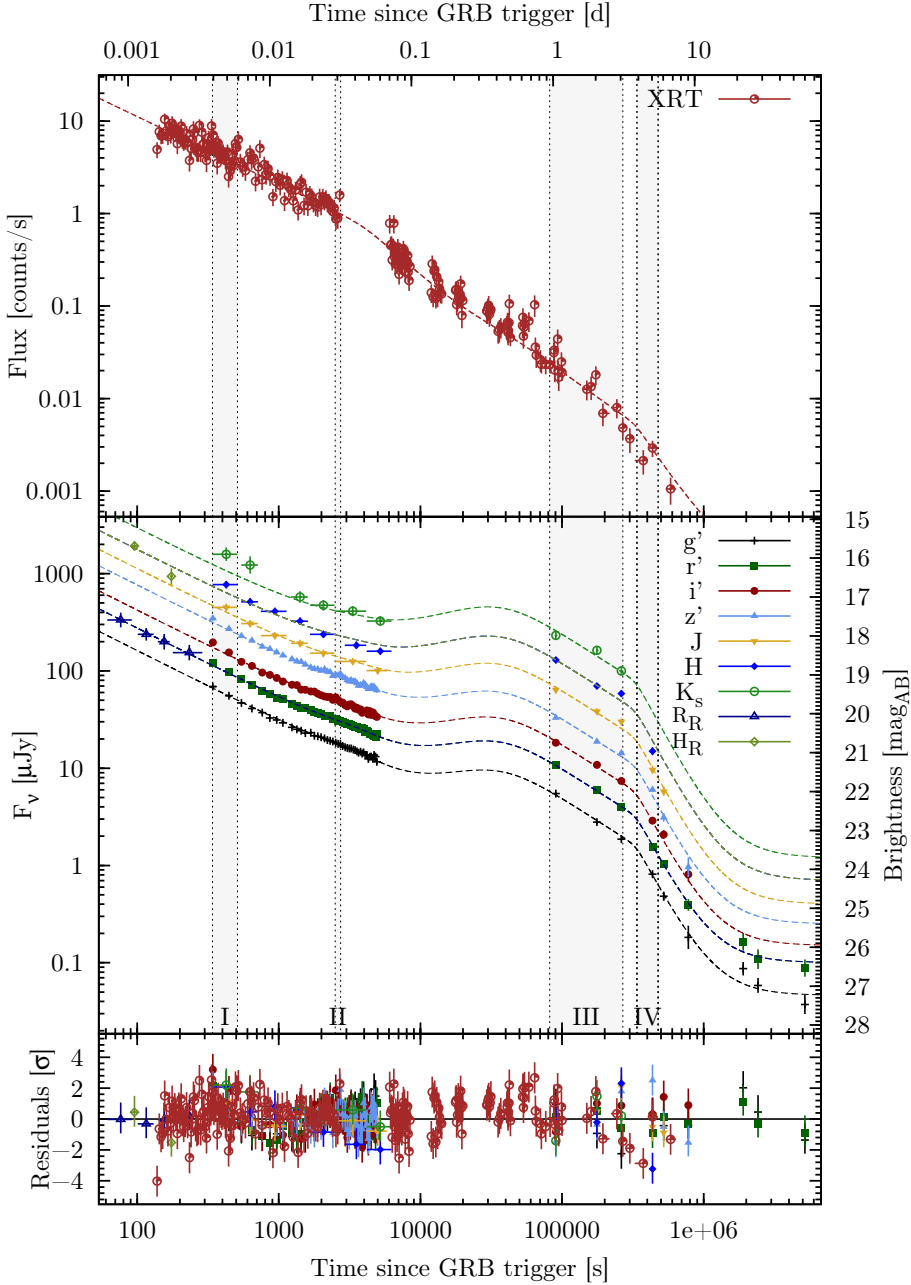}
   \caption{Light curve of the X-ray (top panel) and optical/NIR (middle panel)
     afterglow of GRB 080413B. Bands are offset by $g'+2, r'+1.5, 
         i'+1, z'+0.5, H-0.5, K_s-1$ mag, and REM data $R_R$ and $H_R$ have offsets
       corresponding to GROND data. The bottom panel shows residuals to the combined
       light curve fit. Shown data are corrected for the Galactic foreground extinction and
       transformed into AB magnitudes.
       Upper limits are not shown for better clarity. Gray regions show the time intervals
       where SEDs are reported (Fig. \ref{SED}).}
   \label{lightcurve}
\end{figure*}
 
\subsection{GROND}
	The Gamma-Ray burst Optical Near-infrared Detector (GROND, Greiner et al. \cite{greiner1};
    Greiner et al. \cite{greiner2})
    responded to the \emph{Swift} GRB alert and initiated automated
	observations at 08:56 UT, 5~minutes after the trigger. A predefined sequence
	of observations with successively increasing exposure times was executed
	and images were acquired in the seven photometric bands ($g'r'i'z'JHK_s$) simultaneously.
	The observations continued for two months, and the last of ten epochs was acquired
	on June 11th, 2008. In total, 191 CCD optical individual frames in each 
    $g'r'i'z'$
    and 2718 NIR images of 10~s exposures in $JHK_s$ were obtained.
    The CCD integration time scaled from 45 to 360~s according to the brightness
    of the optical afterglow. 

    A variable point source was detected in all bands 
    (Kr\"uhler et al. \cite{kruhler1})
    by the automated GROND pipeline (K{\" u}pc{\" u} Yolda{\c s} et al. \cite{yoldas}).
    The position of the transient 
    was calculated to be R.A. (J2000) = 21:44:34.67 and Dec (J2000) 
    = $-$19:58:52.4 compared to USNO-B reference field stars (Monet et al. 
    \cite{monet}) with an 
    astrometric uncertainty of $0.\!\!^{\prime\prime}3$. The afterglow was also observed and detected 
    by the Faulkes Telescope South (Gomboc et al. \cite{gomboc}) and Skynet/PROMPT (Brennan et al. \cite{brennan}), 
    and spectroscopy was obtained with the GMOS spectrograph on Gemini-South (Cucchiara et al. 
    \cite{cucchiara}) and FORS1
    on VLT (Vreeswijk et al. \cite{vreeswijk}), both determining a redshift of $1.10$.
    
    The optical and NIR image reduction and photometry were performed using standard
    IRAF tasks (Tody \cite{tody}) similar to the procedure described in detail
    in Kr\"uhler et al. (\cite{kruhler2}). A general model for the point-spread function 
    (PSF) of each image was constructed using bright field stars and fitted to the 
    afterglow. In addition, aperture photometry was carried out, and the results
    were consistent with the reported PSF photometry. All data were corrected
    for a Galactic foreground reddening of $E_{\mathrm{B-V}}=0.04$ mag in the direction 
    of the burst (Schlegel et al. \cite{schlegel}), corresponding to an extinction
    of $A_V=0.11$ using $R_V=3.1$, and in the case of $JHK_s$ data, transformed to AB magnitudes.     

    Optical photometric calibration was performed relative to the magnitudes of six secondary 
    standards in the GRB field, shown in Fig. \ref{chart} and Table \ref{standards}. During 
    photometric
    conditions, a spectrophotometric standard star, SA112-223, a primary
    SDSS standard (Smith et al. \cite{smith}), was observed within a few minutes of observations
    of the GRB field. The obtained zeropoints were corrected
    for atmospheric extinction and used to calibrate stars in the GRB 
    field. The apparent magnitudes 
    of the afterglow were measured with respect to the secondary standards
    reported in Table \ref{standards}. The absolute calibration of $JHK_s$ bands
    was obtained with respect to magnitudes of the Two Micron All Sky Survey
    (2MASS) stars within the GRB field obtained from the 2MASS catalog 
    (Skrutskie et al. \cite{skrutskie}).

\begin{figure*}[ht]
   \centering
   \sidecaption
   \includegraphics[angle=270,width=12cm,clip]{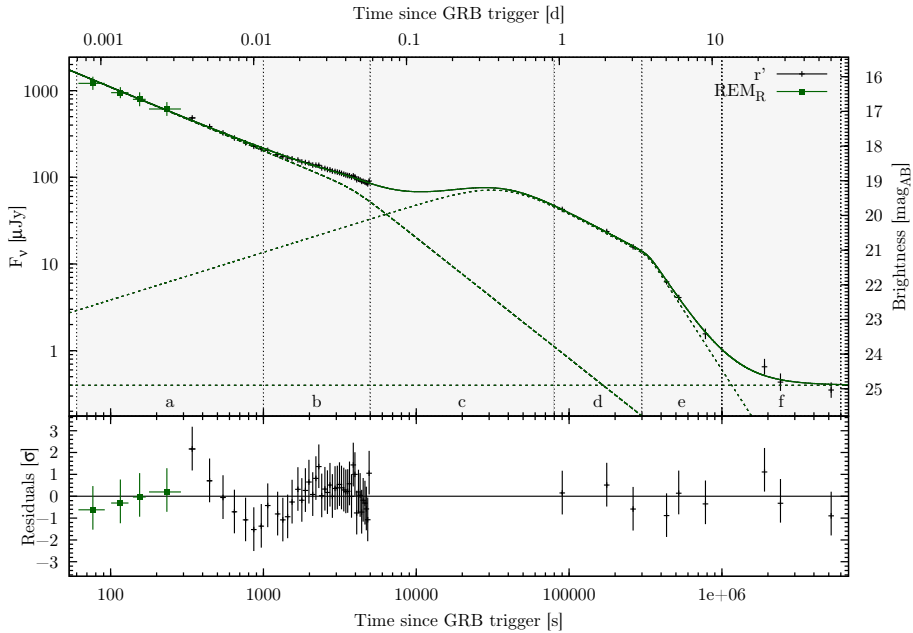}   
   \caption{Three-component fit for GRB 080413B as the superposition of the 
            afterglow emission from the two jets and the host. Shown are the GROND $r'$
            and REM $R$ band data, with all other bands omitted to enhance
            clarity. The additional systematic structure in the residuals between 
            300~s and 5000~s could be additional small-scale variability, which we ignore here.}
   \label{model}
\end{figure*} 
 
\section{Results}
 \subsection{Afterglow light curve}
    The optical/NIR light curve (Fig. \ref{lightcurve}) of the afterglow of GRB
    080413B shows an initial decay with a temporal slope $\alpha = -0.73\pm0.01$,
    followed by a flattening starting at roughly 1~ks. Despite the lack of data
    between 5 and 90~ks, a comparable brightness at the beginning and at the end 
    of the gap (chromatic fading from $\sim0.8$~mag in 
    the $g'$ band to $\sim0.2$~mag in the $K_s$ band) suggests a plateau.
    The light curve then resumes the decay with a steeper temporal slope of 
    $\alpha = -0.95\pm0.02$ until an achromatic break at roughly 330~ks. Owing 
    the achromacity, time, and sharp steepening of the decay, we assume this to be 
    a jet break.
    After this break the afterglow fades with a steep decay of $\alpha = -2.75\pm0.16$. 
    The flattening at the end ($> T_0 + 1$~Ms) of the light curve suggests a faint host galaxy.
    
    The X-ray light curve shows a different evolution. The initial decay has 
    the same temporal slope as the optical/NIR light curve, but the later plateau 
    phase is missing completely. The time of the break at $\sim330$~ks and the 
    decay index after this break is adopted from the optical/NIR data as 
    the X-ray flux does not provide strong constraints in this part of the 
    X- ray light curve. 
    
    Both light curves were jointly fitted with an empirical model consisting of three components
    (see Fig \ref{model}). 
    The first component is composed of two smoothly connected power-laws. The second component
    was needed to model the later rebrightening and uses three smoothly connected
    power-laws. The flattening in the latest part was modeled with a constant flux.
    As a result of the high accuracy of the 
    data and good sampling in the time domain, most parameters were left free to vary and
    are presented in Table \ref{fit}. 

    The only fixed parameters were the smoothnesses $s$
    of all breaks connecting the power-laws and the flux of the host galaxy
    in filters without a detection in the latest flattening phase 
    ($i', z', J, H,$ and $K_s$). The smoothness was fixed to a value
    of $s=10$ in two cases where the power-law decay was steepening in order to be consistent
    with the smoothness of a jet break (Zeh et al. \cite{zeh}) and to a value
    of $s=2$ in the place of the peak of the second component. The flux of the host
    was fixed to values that assume an achromatic afterglow evolution, though this
    is probably not quite correct, as the host is expected to have different colors
    than the afterglow.
    
    The optical/NIR light curve (Fig. \ref{lightcurve}) of the afterglow of GRB
    080413B can be divided into six segments $a,b,c,d,e,$ and $f$, based on the temporal indices shown
    in Fig. \ref{model}. We assign segment $a$ to the first, and segments $c,d,e$
    to the second component. Segment $a$ is the prompt decay dominated by the first component. 
    In segment $b$ we
    see the rising influence of the second component, which then dominates the rest 
    of the later optical light curve and peaks in the third segment $c$. The best fit in the 
    segment $c$ is a plateau-like evolution without any sharp flares. Though we have no data points
    in this segment, magnitudes from Gomboc et al. (\cite{gomboc}) are in good agreement with 
    this interpretation. Segments $d$ and $e$ are fully dominated by the second component 
    with segments $e$ and $f$ showing the rising influence of the constant
    flux, which we interpret as the host galaxy. This host galaxy was detected in the $g'$ and $r'$
    bands, but the stellar mass is not constrained by the optical identification obtained by GROND
    because observations probe the rest-frame wavelengths below the 4000 \AA break, 
    where the mass-to-light ratio can vary by a factor of more than 100.
    The X-ray light curve shows a significantly different evolution, 
    mainly due to a much lower
    contribution from the second component to the total flux. The absence of the 
    rebrightening part gives evidence of the flux from the second component being stronger
    in optical wavelengths and nearly negligible in X-rays. This suggests a different
    physical origin for each component.

\begin{figure*}[ht]
   \sidecaption
   \includegraphics[angle=270,width=12cm,clip]{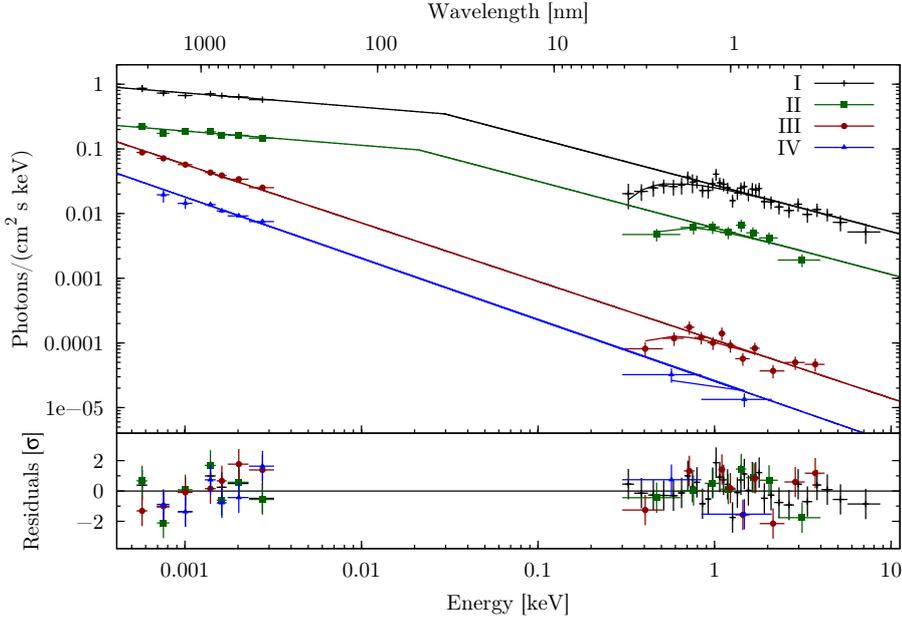}
   \caption{Broad-band spectral energy distribution using the X-ray and optical/NIR data
                   at four epochs indicated in Fig. \ref{lightcurve}. The data were fitted 
                   using a power-law, modified by a Galactic and intrinsic hydrogen
                   column.}
              \label{SED}
\end{figure*}     
  
\subsection{Broad-band spectrum}
    The afterglow spectrum can be parameterized over a broad wavelength range
    using X-ray, optical, and NIR data. Broad-band spectral energy distributions (SED,
    Fig. \ref{SED})
    were constructed at four different time intervals, which are indicated 
    in the light curve (Fig. \ref{lightcurve}).
    Fit parameters of these SEDs are presented in Table \ref{SEDtable}.
    
    As already evident from the lack of the plateau phase in the X-ray light curve, 
    there is a spectral evolution
    between data from the time intervals II and III. The two last optical/NIR 
    SEDs (III and IV) are consistent with a power-law with a spectral index consistent
    with the X-ray spectral index without
    strong signatures of any curvature. There is also no evidence of any spectral
    evolution between SEDs before and after the break at $330$~ks, providing
    more evidence of an achromatic jet break. Both the optical/NIR and X-ray
    emission in these two latest phases probe the same segment 
    of the afterglow synchrotron spectrum with a spectral slope $\beta \sim 0.92$.  
       
    SEDs of phases I and II show evidence of a synchrotron cooling break between 
    the X-ray and optical/NIR frequencies. We fixed the difference in values between 
    optical/NIR and X-ray spectral indices to 0.5 (as predicted by the standard 
    fireball model; Sari et al. \cite{sari4}) but the values come from the fitting. This 
    produced a cooling break that showed a slight drift to lower
    frequencies with time. However, the error on the cooling break frequency is too large
    to claim any trend, and the cooling break is therefore consistent with being constant
    as well. 
    We note that this possible cooling break passage through
    the optical/NIR bands cannot be the cause of rebrightening as it would have 
    the opposite effect, i.e. a steepening of the decay (Sari et al. \cite{sari4}).
 
\subsection{Closure relations}       
    The optical/NIR temporal index $\alpha = -0.73\pm0.01$ of the segment $a$ is consistent 
    within 3$\sigma$ errors with the closure relations 
    (Granot \& Sari \cite{granot3}; Dai \& Cheng \cite{dai2}; Zhang \& M\'esz\'aros \cite{zhang3}; 
    Racusin et al. \cite{racusin})
    for a normal decay in the $\nu_\mathrm{m} < \nu < \nu_\mathrm{c}$ regime, where the jet is interacting
    with a homogeneous ISM and is in the slow cooling phase. The corresponding power-law index
    of electron energy distribution $p = 1.44 \pm 0.16$ is very hard. The X-ray temporal and spectral slopes
    in the segment $a$ are inconsistent with any closure relations. While the spectral slope is different
    from that in the optical/NIR wavelengths, the temporal slopes are similar. 
    
    The temporal index $\alpha = -1.39\pm0.05$ of the second power-law of the first component after the  
    break is within 1$\sigma$ consistent with the closure relations for a post-jet break decay in the 
    $\nu_\mathrm{m} < \nu < \nu_\mathrm{c}$ regime, where the jet is interacting
    with a homogeneous ISM and does not spread. The X-ray slopes are again inconsistent
    with any closure relations.
    
    The initial decay of the second component in segment $d$ with an index of $-0.95\pm0.02$
    is consistent (within 1$\sigma$) with the closure relations for the
    normal (pre-jet break) decay in the $\nu > \nu_c$ regime for a homogeneous ISM and slow cooling
    case. The corresponding electron energy distribution index $p = 1.84 \pm 0.32$ is still
    rather hard but closer to the values typical of GRBs. 
    The late temporal decay is not constrained well by the data, but it is relatively 
    steep and consistent with being achromatic. Fixing the break to be rather sharp 
    (Zeh et al. \cite{zeh}) results in a decay with a temporal index of $\alpha \sim -2.8$, 
    indicative of a post jet break evolution and a break time of roughly 
    330 ks. This light curve slope, however, is not consistent with any 
    closure relation, which might be, at least partially, the result of the 
    parameter fixing in the light curve fitting.

\section{Discussion}
    Rebrightenings of the afterglow light curves are generally associated with
    density inhomogeneities in the circumburst medium or with different forms
    of late energy injections. In this
    section we discuss various possible models for interpreting the optical/NIR rebrightening
    and conclude that the data require the two-component jet model to explain 
    all the light curve features. 
    
 \subsection{Reverse shock emission}
    When the relativistic shell of the fireball ejecta encounters the interstellar
    medium, the reverse shock propagates back into the shocked material and 
    can produce a bright optical flash (Sari \& Piran \cite{sari2}; M\'esz\'aros 
    \& Rees \cite{meszaros}). This emission peaks very early, before the emission
    from the forward shock, and has a steep temporal decay index $\alpha \sim -2$ 
    (Zhang et al. \cite{zhang}; Kobayashi \& Zhang \cite{kobayashi}).
    The reverse shock is therefore inconsistent with being
    the source of the late emission during the plateau phase in segment $c$ 
    since this emission
    peaks at several hours after the burst. A second scenario would be that the 
    initial light curve emission in segment $a$ was the reverse shock component decay 
    and the later plateau 
    was the result of the forward shock emission reaching a peak followed by its slow
    decay, which would then dominate 
    the later light curve (segments $c, d, e$). However, the shallow temporal index during the initial
    decay in segment $a$ is incompatible with emission from a reverse shock. The light curve of the 
    afterglow is therefore incompatible with emission from reverse shocks.
    
 \subsection{Refreshed shock emission}
    Refreshed shocks are produced when slower shells with a lower Lorentz factor catch up 
    with the afterglow shock at
    late times (Rees \& M\'esz\'aros \cite{rees}; Panaitescu \cite{panaitescu2}; 
    Sari \& M\'esz\'aros \cite{sari3};
    Panaitescu et al.\cite{panaitescu}). 
    Each collision then causes a rebrightening in the afterglow light curve.
    After the rebrightening, the afterglow resumes its original decay slope (Granot et al. 
    \cite{granot2}). However, these rebrightenings are generally achromatic (Kumar 
    \& Piran \cite{kumar}) as the slow shell reenergizes the forward shock, which is 
    responsible for both X-ray and optical emission. 
    Therefore, a refreshed shock could not create the chromatic rebrightening
    in the light curve of the afterglow after the initial decay. Different temporal
    indices before and after this event moreover exclude refreshed shocks as a
    feasible explanation for the evolution of the light curve.
    
 \subsection{Inhomogeneous density profile of the ISM}
    Variations in the external density provide a possible explanation
    for the temporal variability of the GRB afterglow light curves within
    the external shock framework (Lazzati et al. \cite{lazzati}; Nakar et al. 
    \cite{nakar2}; Zhang et al. \cite{zhang2}; Nakar \& Piran \cite{nakar3}; 
    Ioka et al. \cite{ioka}; Wang \& Loeb \cite{wang}; Dai \& Lu \cite{dai}).
    Such variations might be the result of the interstellar medium turbulence
    or variability in the winds from the progenitor. The first case might be
    characterized by either an abrupt change in the radial density or
    density clumps on top of a smooth background. The latter can be a case
    of the wind termination shock, which is an abrupt increase in the radial density 
    between wind environments of two evolutionary stages of the massive 
    progenitor. Models suggest that these inhomogeneities will have 
    a clear observational signature in the form of the optical afterglow
    light curve rebrightening. 
    
    The initial decay of the light curve of GRB 080413B is a smooth power-law
    with $\alpha\sim -0.7$. At $T\sim 0.12$ d $\alpha$ becomes 
    positive over a factor $\sim 2-3$ in time. According to Nakar \& Granot 
    (\cite{nakar}) and van Eerten et al. (\cite{eerten}), 
    such a large increase in $\alpha$ over a relatively small
    factor in time is not expected from variations in the external density. 
    Even though our temporal index during the rebrightening is based on the 
    fit alone, real data before and after this gap show that a smooth power
    law connected with very sharp breaks and with a temporal index very near 
    zero would be needed 
    to connect these data points without a rebrightening with a peak. While the wind
    termination shock could explain the lack of the rebrightening feature in 
    X-ray band, the expected increase in temporal index above the cooling frequency
    is too small to be compatible with our optical light curve. 
 
 \subsection{Two-component jet}
    The generic two-component jet model consists of a narrow and highly relativistic
    jet, responsible for the prompt afterglow emission, and of a wider and moderately
    relativistic jet, dominant in the later afterglow emission (Peng et al. \cite{peng}; 
    Granot et al. \cite{granot}; Berger et al. \cite{berger}; Racusin et al. \cite{racusin2}). 
    For an on-axis geometry, the resulting afterglow 
    light curve is a superposition of these two components, where the decelerating
    narrow jet creates the initial decay and the wide jet dominates the later emission
    that rises during the pre-deceleration phase, followed by the shallow decay with 
    a possible jet break. The relative energies and jet structure then define the
    light curve morphology. 
    
    The light curve is well-fitted (red. $\chi^2 = 1.08$) with the sum of the two components
    that we relate to the two afterglow
    jets, where both are viewed on-axis and are coaxial. The initial shallow decay
    phase of segment $a$ could be the result of the emission of the decelerating narrow jet. Given 
    that we do not see any rising part in the early light curve and that
    even the very early data from the REM telescope have the same decay slope as the later
    GROND data, we can safely assume that we see the narrow jet on-axis. 
    From the time of the jet break at around $3.9$~ks, we can calculate the opening
    angle (Sari et al. \cite{sari}) of the narrow jet as $\theta_n \sim 1.7^\circ$,
    substituting the measured quantities and normalizing to the typical values
    $n = 1$~cm$^{-3}$ and $\eta = 0.2$ (Bloom et al. \cite{bloom}). Assuming the time
    of the first $R'$ band data point to be upper limit on the time of the emission
    peak, we calculate the 
    initial Lorentz factor (Molinari et al. \cite{molinari}) to be $\Gamma_n > 188$.
    These values lead to the beaming factor and the true gamma-ray energy release (Frail et al. 
    \cite{frail}; Bloom et al. \cite{bloom}) of $f_b = (1-\cos \theta_{jet}) = 4.4 \times 10^{-4}$
    and $E_{\gamma,n} = 7.9 \times 10^{48}$~erg.
    
    The wide-jet component might be responsible for the rebrightening starting
    at around 10~ks. However, this second component is visible
    even earlier in the initial decay phase, where the light curve gets shallower 
    (segment $b$). The initial rising temporal index is compatible with the jet decelerating
    in the circumburst medium. The wide jet
    is therefore seen on-axis as well, and both jets can be considered coaxial.
    The jet break at roughly 330~ks indicates an opening angle of the wide jet
    of $\theta_w \sim 9^\circ$. The initial Lorentz factor, corresponding
    to the peak of the second jet at 37~ks, is then $\Gamma_w \sim 18.5$.
        
\section{Conclusions}
    In this paper we study the optical/NIR light curve produced by the afterglow of GRB 080413B. 
    The possibility that the jet of this GRB might have a narrow ultra-relativistic
    core and a wider, mildly relativistic outer component has been indicated by 
    the observation of the afterglow emission.    
    An on-axis coaxial two-component jet model provides a consistent description
    of the properties of GRB 080413B, and can additionally explain the wide range
    of light curve evolutions, the difference between optical/NIR
    and X-ray light curves, and the chromatic evolution of the optical 
    light curve itself.
    
    The comparison with the two most prominent light curves modeled by the 
    two-component jet to date - GRB 050315 and GRB 080319B - reveal consistency
    with the GRB 080413B afterglow light curve. 
    The X-ray light curve of the afterglow of GRB 050315 (Granot et al. \cite{granot};
    Nousek et al. \cite{nousek}) shows a remarkable resemblance to the optical/NIR light curve
    evolution of the afterglow of GRB 080413B. If we neglect the very steep tail of the prompt GRB emission, 
    the initial XRT light curve of GRB 050315 is dominated by the narrow jet, followed
    by a slight rebrightening at around 1.5~ks caused by the wide jet in its pre-deceleration
    phase. After the peak, the light curve decay is dominated by the emission from the wide
    jet. Times of jet breaks of narrow ($\sim9$~ks) and wide ($\sim200$~ks) components, 
    as well as their opening angles $\theta_w = 2 \theta_n = 3.2^\circ$ 
    (Granot et al. \cite{granot}), are within an order comparable with those of GRB 080413B. 
    
    The X-ray light curve of the naked-eye GRB 080319B (Racusin et al. \cite{racusin2})
    shows similar evolution. The narrow jet dominates the first $\sim40$~ks of the afterglow.
    After the narrow jet decays, the wide jet dominates the late afterglow. There is no
    rising part of the wide jet and thus no sharp rebrightening or plateau, so the wide jet merely
    makes the decay flatter. The optical light curve is missing the emission from the narrow
    jet, suggesting that the optical flux from the wide jet must be much stronger than that of the
    narrow jet. The jet break of the narrow jet at $\sim2.8$~ks, which corresponds to an extremely 
    narrow opening angle of $0.2^\circ$, is the earliest of these three bursts.
    The jet break of the wide component with opening angle $\sim4^\circ$ is, on the other hand,
    the latest at roughly 1~Ms.  
    In general, the X-ray light curves of GRBs 050315, 080319B and the optical light curve of GRB 080413B are 
    very similar. However, the afterglow of GRB 080413B is the only one showing both components
    in the optical/NIR wavelengths, while the emission from the wide jet in the X-rays is negligible. 
    The X-ray flux from the wide jet must then be much less prominent than for the narrow jet.
    
    Following this line of reasoning the relative fluxes in optical/NIR and X-ray
    of the narrow and wide jets can be explained in the following way.
    The SED of the narrow jet (intervals I and II) shows a break, while that
    of the wider jet does not. For both jets we have argued that
    we cover the slow cooling regime. The spectral slope of the
    wide component implies that the cooling break is at frequencies below
    the near-infrared bands. Both the cooling frequency and
    the maximum power depend on the product of Lorentz factor
    $\Gamma$ of the shocked fluid and the magnetic field strength. It
    is generally assumed that the narrow jet comes with a larger
    Lorentz factor than the wide one, and a similar assumption can be reasonably
    made about the (self-created) magnetic field.
    The SEDs of the two jets show us that the product $\Gamma *B$
    of the wide jet, and consequently also the emission at X-ray energies,
    are at least a factor 100 less than for the narrow one.
    Therefore, the wide jet does not contribute to the X-ray
    emission in any significant way.
    The situation is different in the optical/NIR since cooling
    break of the narrow jet leads to a reduced flux by a factor of
    $\approx$ 10 relative to a spectrum with no cooling break between the
    optical/NIR and X-rays. Consequently, the optical/NIR emission of the
    wide jet is much more prominent than for the narrow jet.
    
    The values derived from the modeling of GRB 080413B afterglow are fairly consistent with 
    the collapsar jet breakout model of Zhang et al. (\cite{zhang5}), where the numerical 
    simulations predict $\theta_n = 3-5^\circ$, 
    $\Gamma_n \gtrsim 100$ for the narrow component and $\theta_w \sim 10^\circ$, $\Gamma_w \sim 15$ 
    for the wide component (Peng et al. \cite{peng}). The characteristic Lorentz factors 
    are very similar to those of the hydromagnetically accelerated, initially neutron-rich jet model
    of Vlahakis et al. (\cite{vlahakis}), where $\Gamma_n \sim 200$ and $\Gamma_w \sim 15$.
    These two models are distinguished by the ratio of the kinetic energy injected into the two components.
    For values typical of the collapsar model ($E_w / E_n \sim 0.1$), Peng et al. 
    (\cite{peng}) predict that the contribution of the narrow component dominates at all times. 
    However, for $E_w \gtrsim 2 E_n$ (as in the 
    neutron-rich hydromagnetic model), the narrow component dominates at early times but the contribution 
    of the wide jet becomes dominant around the deceleration time of the wide jet. 
    If $E_w > E_n$, the jet break 
    of the narrow jet could be masked by the rise (and subsequent dominance) of the flux from the wide jet 
    as the deceleration time of the wide component is approached. That the only visible 
    jet break in the optical light curve is the one of the wide jet may lead to overestimating
    the emitted gamma-ray energy if the opening angle of the wide jet is used in converting 
    the measured energy into the beaming-corrected energy (see Peng et al \cite{peng} for detailed
    discussion). Because the deceleration time of the wide component is much longer than for 
    the narrow component, a bump is expected to show up in the decaying light curve of the narrow 
    component owing the emission of the wide component at its deceleration time. These predictions are 
    in perfect agreement with our data, suggesting that the two-component jet model can be placed
    among models that explain the variability in the early light curves of the GRB afterglows.

\begin{acknowledgements}
TK acknowledges support by the DFG cluster of excellence Origin and 
Structure of the Universe.\\
Part of the funding for GROND (both hardware and 
personnel) was generously granted from the Leibniz-Prize to Prof. G. 
Hasinger (DFG grant HA 1850/28-1).\\
SK and ARossi acknowledge support by DFG grant Kl 766/11-3 and ARossi
additionally from the BLANCEFLOR Boncompagni- Ludovisi, n\'ee Bildt
foundation.\\
MN acknowledges support by DFG grant SA 2001/2-1.\\
FOE acknowledges support by the DAAD.\\
Swift/XRT: This work made use of data supplied by the UK Swift Science 
Data Centre at the University of Leicester.\\
We thank the anonymous referee for constructive comments that helped to improve the paper.
\end{acknowledgements}

\newpage
\begin{table*}
\caption{$g'r'i'z'$ photometric data}             
\label{griz}      
\centering                         
\begin{tabular}{r c c c c c}        
\hline     
\\            
$T_\mathrm{mid} - T_0$ [ks] & Exposure [s] & \multicolumn{4}{c}{
Brightness$^{\mathrm{(a)}}$ mag$_\mathrm{AB}$} \\
\hline    
\\
 & & $g'$ & $r'$ & $i'$ & $z'$ \\
\hline                                   
0.0765 & 30 & & $16.19 \pm 0.19$ \\ 
0.1158 & 30 & & $16.46 \pm 0.16$ \\
0.1549 & 30 & & $16.65 \pm 0.20$ \\
0.2334 & 110 & & $16.93 \pm 0.20$ \\
0.3421 & 35	& $	17.44	\pm	0.05	$ & $	17.28	\pm	0.03	$ & $	17.24	\pm	0.04	$ & $	17.10	\pm	0.05	$ \\
0.4441 & 35	& $	17.67	\pm	0.04	$ & $	17.53	\pm	0.03	$ & $	17.49	\pm	0.04	$ & $	17.37	\pm	0.04	$ \\
0.5443 & 35 & $	17.86	\pm	0.04	$ & $	17.71	\pm	0.03	$ & $	17.73	\pm	0.03	$ & $	17.56	\pm	0.04	$ \\
0.6463 & 35	& $	18.00	\pm	0.04	$ & $	17.86	\pm	0.03	$ & $	17.84	\pm	0.04	$ & $	17.67	\pm	0.04	$ \\
0.7646 & 35	& $	18.11	\pm	0.04	$ & $	18.00	\pm	0.04	$ & $	18.01	\pm	0.04	$ & $	17.83	\pm	0.05	$ \\
0.8657 & 35 & $	18.24	\pm	0.04	$ & $	18.10	\pm	0.03	$ & $	18.07	\pm	0.04	$ & $	17.89	\pm	0.04	$ \\
0.9677 & 35 & $	18.29	\pm	0.03	$ & $	18.17	\pm	0.03	$ & $	18.15	\pm	0.04	$ & $	17.97	\pm	0.04	$ \\
1.0679 & 35	& $	18.37	\pm	0.04	$ & $	18.21	\pm	0.03	$ & $	18.24	\pm	0.04	$ & $	18.05	\pm	0.04	$ \\
1.2390 & 35	& $	18.49	\pm	0.04	$ & $	18.33	\pm	0.04	$ & $	18.32	\pm	0.04	$ & $	18.15	\pm	0.05	$ \\
1.3401 & 35	& $	18.54	\pm	0.04	$ & $	18.39	\pm	0.03	$ & $	18.33	\pm	0.04	$ & $	18.21	\pm	0.04	$ \\
1.4394 & 35	& $	18.57	\pm	0.04	$ & $	18.43	\pm	0.03	$ & $	18.45	\pm	0.04	$ & $	18.22	\pm	0.04	$ \\
1.5396 & 35	& $	18.63	\pm	0.04	$ & $	18.45	\pm	0.03	$ & $	18.46	\pm	0.04	$ & $	18.29	\pm	0.04	$ \\
1.6813 & 35	& $	18.63	\pm	0.04	$ & $	18.49	\pm	0.04	$ & $	18.49	\pm	0.05	$ & $	18.36	\pm	0.05	$ \\
1.7824 & 35	& $	18.70	\pm	0.03	$ & $	18.54	\pm	0.03	$ & $	18.50	\pm	0.04	$ & $	18.40	\pm	0.04	$ \\
1.8835 & 35	& $	18.72	\pm	0.04	$ & $	18.56	\pm	0.03	$ & $	18.56	\pm	0.04	$ & $	18.41	\pm	0.04	$ \\
1.9863 & 35	& $	18.74	\pm	0.03	$ & $	18.58	\pm	0.03	$ & $	18.59	\pm	0.04	$ & $	18.45	\pm	0.04	$ \\
2.1021 & 35	& $	18.77	\pm	0.04	$ & $	18.64	\pm	0.04	$ & $	18.62	\pm	0.04	$ & $	18.41	\pm	0.05	$ \\
2.2041 & 35	& $	18.81	\pm	0.04	$ & $	18.64	\pm	0.03	$ & $	18.64	\pm	0.04	$ & $	18.47	\pm	0.04	$ \\
2.3034 & 35	& $	18.83	\pm	0.04	$ & $	18.64	\pm	0.04	$ & $	18.64	\pm	0.04	$ & $	18.48	\pm	0.04	$ \\
2.4062 & 35	& $	18.85	\pm	0.04	$ & $	18.72	\pm	0.04	$ & $	18.71	\pm	0.04	$ & $	18.58	\pm	0.04	$ \\
2.5237 & 35	& $	18.88	\pm	0.04	$ & $	18.73	\pm	0.04	$ & $	18.65	\pm	0.04	$ & $	18.56	\pm	0.06	$ \\
2.6240 & 35	& $	18.91	\pm	0.04	$ & $	18.76	\pm	0.03	$ & $	18.72	\pm	0.04	$ & $	18.57	\pm	0.04	$ \\
2.7242 & 35	& $	18.91	\pm	0.04	$ & $	18.77	\pm	0.03	$ & $	18.77	\pm	0.04	$ & $	18.52	\pm	0.04	$ \\
2.8261 & 35	& $	18.96	\pm	0.04	$ & $	18.81	\pm	0.04	$ & $	18.78	\pm	0.04	$ & $	18.61	\pm	0.04	$ \\
2.9428 & 35	& $	19.00	\pm	0.04	$ & $	18.82	\pm	0.04	$ & $	18.85	\pm	0.05	$ & $	18.64	\pm	0.05	$ \\
3.0439 & 35	& $	19.01	\pm	0.03	$ & $	18.84	\pm	0.03	$ & $	18.88	\pm	0.04	$ & $	18.69	\pm	0.04	$ \\
3.1441 & 35	& $	19.03	\pm	0.03	$ & $	18.85	\pm	0.03	$ & $	18.83	\pm	0.04	$ & $	18.72	\pm	0.05	$ \\
3.2443 & 35	& $	19.03	\pm	0.04	$ & $	18.88	\pm	0.04	$ & $	18.83	\pm	0.04	$ & $	18.69	\pm	0.05	$ \\
3.3566 & 35	& $	19.06	\pm	0.04	$ & $	18.90	\pm	0.04	$ & $	18.87	\pm	0.05	$ & $	18.73	\pm	0.06	$ \\
3.4577 & 35	& $	19.07	\pm	0.04	$ & $	18.92	\pm	0.03	$ & $	18.94	\pm	0.04	$ & $	18.72	\pm	0.05	$ \\
3.5580 & 35	& $	19.10	\pm	0.04	$ & $	18.94	\pm	0.03	$ & $	18.91	\pm	0.04	$ & $	18.77	\pm	0.05	$ \\
3.6599 & 35	& $	19.14	\pm	0.04	$ & $	18.94	\pm	0.04	$ & $	18.97	\pm	0.04	$ & $	18.83	\pm	0.05	$ \\
3.7731 & 35	& $	19.11	\pm	0.05	$ & $	18.98	\pm	0.04	$ & $	18.96	\pm	0.05	$ & $	18.81	\pm	0.06	$ \\
3.8733	& 35 & $ 19.16	\pm	0.05	$ & $	18.95	\pm	0.04	$ & $	19.05	\pm	0.05	$ & $	18.83	\pm	0.05	$ \\
3.9735	& 35 & $ 19.13	\pm	0.04	$ & $	18.98	\pm	0.04	$ & $	19.03	\pm	0.04	$ & $	18.77	\pm	0.04	$ \\
4.0764	& 35 & $ 19.19	\pm	0.04	$ & $	19.06	\pm	0.04	$ & $	19.03	\pm	0.05	$ & $	18.81	\pm	0.05	$ \\
4.1913	& 35 & $ 19.20	\pm	0.05	$ & $	19.04	\pm	0.04	$ & $	18.98	\pm	0.05	$ & $	18.87	\pm	0.06	$ \\
4.2924	& 35 & $ 19.30	\pm	0.04	$ & $	19.06	\pm	0.04	$ & $	19.08	\pm	0.04	$ & $	18.92	\pm	0.05	$ \\
4.3943	& 35 & $ 19.24	\pm	0.05	$ & $	19.11	\pm	0.04	$ & $	19.00	\pm	0.04	$ & $	18.90	\pm	0.05	$ \\
4.4963	& 35 & $ 19.24	\pm	0.04	$ & $	19.10	\pm	0.04	$ & $	19.10	\pm	0.05	$ & $	18.85	\pm	0.04	$ \\
4.6172	& 35 & $ 19.29	\pm	0.09	$ & $	19.13	\pm	0.05	$ & $	19.03	\pm	0.06	$ & $	18.86	\pm	0.08	$ \\
4.7148	& 35 & $ 19.20	\pm	0.06	$ & $	19.15	\pm	0.04	$ & $	19.09	\pm	0.05	$ & $	18.86	\pm	0.05	$ \\
4.8159	& 35 & $ 19.24	\pm	0.06	$ & $	19.18	\pm	0.04	$ & $	19.13	\pm	0.06	$ & $	18.92	\pm	0.07	$ \\
4.9179	& 35 & $ 19.37	\pm	0.07	$ & $	19.10	\pm	0.05	$ & $	19.15	\pm	0.08	$ & $	18.95	\pm	0.06	$ \\
90.3010	& 2733 & $ 20.19 \pm 0.04	$ & $	19.91	\pm	0.04	$ & $	19.81	\pm	0.05	$ & $	19.65	\pm	0.04	$ \\
176.4193 & 3805	& $	20.92 \pm 0.05	$ & $	20.55	\pm	0.04	$ & $	20.38	\pm	0.05	$ & $	20.27	\pm	0.05	$ \\
262.4098 & 3556	& $	21.36 \pm 0.04	$ & $	21.00	\pm	0.04	$ & $	20.79	\pm	0.04	$ & $	20.57	\pm	0.05	$ \\
434.6533 & 4046	& $	22.26 \pm 0.05	$ & $	22.01	\pm	0.04	$ & $	21.82	\pm	0.06	$ & $	21.51	\pm	0.06	$ \\
522.1878 & 3036	& $	22.83 \pm 0.12	$ & $	22.46	\pm	0.07	$ & $	22.17	\pm	0.11	$ & $	22.22	\pm	0.11	$ \\
780.5056 & 3520	& $	23.88 \pm 0.30	$ & $	23.50	\pm	0.15	$ & $	23.20	\pm	0.18	$ & $	23.49	\pm	0.25	$ \\
1901.2933 & 2989 & $ 24.69 \pm 0.17	$ & $	24.45	\pm	0.23	$ & $	> 23.54	$ & $	> 23.34	$ \\
2418.4846 & 7770 & $ 25.12 \pm 0.19	$ & $	24.90	\pm	0.25	$ & $	> 23.78 $ & $	> 23.36	$ \\
5185.0714 & 6995 & $ 25.61 \pm 0.24	$ & $	25.13	\pm	0.22	$ & $	> 24.22	$ & $	> 23.87	$ \\
\hline
\end{tabular}
\begin{list}{}{}
\item[$^{\mathrm{(a)}}$] Not corrected for Galactic foreground reddening.
\end{list}
\end{table*}

\begin{table*}
\caption{$JHK_s$ photometric data}             
\label{JHK}      
\centering                         
\begin{tabular}{r c c c c}        
\hline     
\\            
$T_\mathrm{mid} - T_0$ [ks] & Exposure [s] & \multicolumn{3}{c}{
Brightness$^{\mathrm{(a)}}$ mag$_\mathrm{AB}^{\mathrm{(b)}}$}  \\
\hline    
\\
 & & $J$ & $H$ & $K_s$  \\
\hline 
0.0956  &   10  & & $ 16.19   \pm 0.12    $ & \\
0.1746  &   10  & & $ 16.97   \pm 0.21    $ & \\
0.4251	&	168	& $	17.30	\pm	0.07	$ & $	17.20	\pm	0.07	$ & $	16.92	\pm	0.15	$ \\
0.6275	&	168	& $	17.71	\pm	0.06	$ & $	17.65	\pm	0.08	$ & $	17.19	\pm	0.20	$ \\
0.9392	&	385	& $	18.03	\pm	0.06	$ & $	17.89	\pm	0.07	$ & $				$ \\
1.4126	&	384	& $	18.22	\pm	0.07	$ & $	18.14	\pm	0.07	$ & $	18.01	\pm	0.12	$ \\
2.0667	&	808	& $	18.47	\pm	0.07	$ & $	18.48	\pm	0.07	$ & $	18.23	\pm	0.11	$ \\
3.3229	&	1637 (1216 for $H$)	& $	18.69	\pm	0.06	$ & $	18.76	\pm	0.07	$ & $	18.38	\pm	0.10	$ \\
4.9974	&	2065 (1650 for $J$)	& $	18.93	\pm	0.06	$ & $	18.91	\pm	0.06	$ & $	18.63	\pm	0.08	$ \\
90.3234	&	2780	& $	19.42	\pm	0.06	$ & $	19.14	\pm	0.05	$ & $	19.00	\pm	0.05	$ \\
176.4426	&	3854	& $	19.97	\pm	0.05	$ & $	19.81	\pm	0.05	$ & $	19.39	\pm	0.05	$ \\
262.4340	&	3605	& $	20.23	\pm	0.05	$ & $	20.00	\pm	0.05	$ & $	19.91	\pm	0.08	$ \\
434.6775	&	4096	& $	21.47	\pm	0.08	$ & $	21.48	\pm	0.08	$ & $	> 20.28			$ \\
522.2111	&	3084	& $	22.03	\pm	0.09	$ & $	> 21.60			$ & $	> 20.56			$ \\
780.5298	&	3567	& $	> 22.16			$ & $	> 21.48			$ & $	> 20.55			$ \\
1901.3193	&	3040	& $	> 21.96			$ & $	> 21.35			$ & $	> 20.57			$ \\
2418.5088	&	7822	& $	> 22.36			$ & $	> 21.63			$ & $	> 20.90			$ \\
5185.0938	&	7041	& $	> 22.57			$ & $	> 21.92			$ & $	> 20.97			$ \\
\hline
\end{tabular}
\begin{list}{}{}
\item[$^{\mathrm{(a)}}$] Not corrected for Galactic foreground reddening. Converted
to AB magnitudes for consistency with Table \ref{griz}.
\item[$^{\mathrm{(a)}}$] For the SED fitting, the additional error of the absolute
calibration of 0.07 ($J$ and $H$) and 0.09 ($K_s$) mag was added.
\end{list}
\end{table*}

\begin{table*}
\caption{Secondary standards in the GRB field in the GROND
filter bands used for the calibration}             
\label{standards}      
\centering                         
\begin{tabular}{c c c c c c c c c}        
\hline\hline                 
Star & R.A., Dec & $g'$ & $r'$ & $i'$ & $z'$ & $J$ & $H$ & $K_s$ \\
 number & [J2000] & (mag$_\mathrm{AB}$) & (mag$_\mathrm{AB}$) & (mag$_\mathrm{AB}$) & (mag$_\mathrm{AB}$) &
 (mag$_\mathrm{Vega}$) & (mag$_\mathrm{Vega}$) & (mag$_\mathrm{Vega}$) \\
\hline                        
1 & 21:44:32.81, $-$19:58:39.4 & $18.05\pm 0.03$ & $17.14 \pm 0.03$ & $16.93 \pm 0.03$ & $16.70 \pm 0.03$ & 
                         $15.75 \pm 0.05$ & $15.20 \pm 0.05$ & $15.03 \pm 0.06$ \\      
2 & 21:44:32.38, $-$19:58:45.1 & $18.90 \pm 0.03$ & $17.33 \pm 0.03$ & $16.08 \pm 0.04$ & $15.40 \pm 0.03$ & 
                         $14.11 \pm 0.05$ & $13.49 \pm 0.05$ & $13.23 \pm 0.06$ \\
3 & 21:44:33.65, $-$19:58:07.7 & $17.25 \pm 0.03$ & $16.59 \pm 0.03$ & $16.47 \pm 0.03$ & $16.29 \pm 0.03$ & 
                         $15.46 \pm 0.05$ & $15.00 \pm 0.05$ & $14.82 \pm 0.06$ \\
4 & 21:44:35.98, $-$19:57:47.9 & $16.41 \pm 0.03$ & $15.45 \pm 0.03$ & $15.15 \pm 0.03$ & $14.88 \pm 0.03$ & 
                         $13.87 \pm 0.05$ & $13.37 \pm 0.05$ & $13.15 \pm 0.06$ \\
5 & 21:44:38.98, $-$19:59:09.2 & $16.26 \pm 0.03$ & $15.49 \pm 0.03$ & $15.28 \pm 0.03$ & $15.08 \pm 0.03$ & 
                         $14.13 \pm 0.05$ & $13.57 \pm 0.05$ & $13.42 \pm 0.06$ \\
6 & 21:44:32.50, $-$19:59:44.2 & $16.10 \pm 0.03$ & $15.34 \pm 0.03$ & $15.19 \pm 0.03$ & $14.99 \pm 0.03$ & 
                         $14.19 \pm 0.06$ & $13.53 \pm 0.05$ & $13.52 \pm 0.06$ \\
\hline                                   
\end{tabular}
\end{table*}

\begin{table*}
\caption{Light curve fit with smoothness, break-time, and power-law slope parameters for both components}
\label{fit}      
\centering                         
\begin{tabular}{c c c c c c c c c c c c}        
\hline     
\\            
$ F_\nu(t)$ & $\alpha_1$ & $t_1 [s]$ & $s_1$ & $\alpha_2$ & $t_2 [s]$ & $s_2$ &
$\alpha_3$ & $\chi^2$/d.o.f.\\    
\hline 
DPL$^{\mathrm{(a)}}$ & $-0.73 \pm 0.01$ & $3903 \pm 181$ & 10 & $-1.39 \pm 0.05$ & & & & \multirow{2}{*}{462 / 429} \\
TPL$^{\mathrm{(b)}}$ & $0.55 \pm 0.05$ & $36999 \pm 2761$ & 2 & $-0.95 \pm 0.02$ &
$332437 \pm 11375$ & 10 & $-2.75 \pm 0.16$ \\
\hline
\end{tabular}
\begin{list}{}{}
\item[$^{\mathrm{(a)}}$] Smoothly connected double power-law
\item[$^{\mathrm{(b)}}$] Smoothly connected triple power-law
\end{list}
\end{table*}

\begin{table*}
\caption{SED fits of four epochs using X-ray and optical/NIR data}             
\label{SEDtable}      
\centering                         
\begin{tabular}{c c c c c c}        
\hline     
\\            
Epoch & Optical/NIR spectral index $\beta$ & X-ray spectral index $\beta$ & 
Cooling frequency [eV] & N$_H^{\mathrm{(a)}}$ [$10^{22}$ cm$^{-2}$] & $\chi^2$/d.o.f  \\    
\hline 
I & $0.22 \pm 0.08$ & $0.72 \pm 0.08$ & $29.6^{+10.4}_{-24.2}$ & $0.16 \pm 0.12$ & 24/36 \\
II & $0.22 \pm 0.04$ & $0.72 \pm 0.04$ & $21.1^{+26.7}_{-12.0}$ & $0.16$ (frozen) & 16/12 \\
III & \multicolumn{2}{c}{$0.90 \pm 0.05$} & & $0.16$ (frozen) & 23/17 \\
IV & \multicolumn{2}{c}{$0.95 \pm 0.05$} & & $0.16$ (frozen) & 10/6 \\
\hline
\end{tabular}
\begin{list}{}{}
\item[$^{\mathrm{(a)}}$] Intrinsic hydrogen column density, in excess of the frozen 
Galactic foreground of $N_\mathrm{H}=3.1 \times 10^{20}$ cm$^{-2}$.
\end{list}
\end{table*}

\end{document}